\documentclass[pre,aps,twocolumn,amsmath,eqsecnum,unsortedaddress,nofootinbib,10pt]{revtex4}

\usepackage{amssymb}
\usepackage{amsmath}
\usepackage{graphicx}
\usepackage{amsbsy}
\usepackage{color}
\usepackage{verbatim}
\usepackage{enumerate}

\begin{document}
\title{A unified description of solvent effects in the helix-coil transition}

\author{Artem Badasyan}
\email{abadasyan@gmail.com}
\affiliation{Materials Research Laboratory, University of Nova Gorica,\\ Vipavska 13, SI-5000 Nova Gorica, Slovenia, EU}

\author{Shushanik A. Tonoyan}
\affiliation{Department of Molecular Physics, Yerevan State University,\\ A.Manougian Str.1, 375025, Yerevan, Armenia}

\author{Achille Giacometti}
\affiliation{Dipartimento di Scienze Molecolari e Nanosistemi, Universit\`a Ca' Foscari Venezia,
Calle Larga S. Marta DD2137, I-30123 Venezia, Italy, EU}

\author{Rudolf Podgornik}
\affiliation{Department of Theoretical Physics, J. Stefan Institute and Department of Physics, Faculty of Mathematics and Physics, University of Ljubljana - SI-1000 Ljubljana, Slovenia, EU and Department of Physics, University of Massachusetts, Amherst, MA 01003-9337 USA}

\author{V. Adrian Parsegian}
\affiliation{Department of Physics, University of Massachusetts, Amherst, MA 01003-9337 USA}

\author{Yevgeni Sh. Mamasakhlisov and Vladimir F. Morozov}
\affiliation{Department of Molecular Physics, Yerevan State University,\\ A. Manougian Str.1, 375025, Yerevan, Armenia}
\date{\today}

\begin{abstract}

We analyze the problem of the helix-coil transition in explicit solvents analytically by using spin-based models incorporating two different mechanisms of solvent action: explicit solvent action through the formation of solvent-polymer hydrogen bonds that can compete with the intrinsic intra-polymer hydrogen bonded configurations (competing interactions) and implicit solvent action, where the solvent-polymer interactions tune biopolymer configurations by changing the activity of the solvent (non-competing interactions). The overall spin Hamiltonian is comprised of three terms: the background \emph{in vacuo} Hamiltonian of the "Generalized Model of Polypeptide Chain" type and two additive terms that  account for the two above mechanisms of solvent action. We show that on this level the solvent degrees of freedom can be {\sl  explicitly} and {\sl exactly} traced over, the ensuing effective partition function combining all the solvent effects in a unified framework. In this way we are able to address helix-coil transitions for polypeptides, proteins, and DNA, with different buffers and different external constraints. Our spin-based effective Hamiltonian is applicable for treatment of such diverse phenomena as cold denaturation, effects of osmotic pressure on the cold and warm denaturation, complicated temperature dependence of the hydrophobic effect as well as providing a conceptual base for understanding the behavior of Intrinsically Disordered Proteins and their analogues.

\end{abstract}


\maketitle
\section{Introduction}
\label{sec:introduction}

The inside of biological cells is a crowded and complex environment composed of chemically different low- and high-molecular weight compounds dissolved in the aqueous solvent. In order to model this internal cellular milieu \emph{in vitro}, biopolymers, such as polypeptides, proteins, and DNA, are usually studied in aqueous solutions of different composition and with different imposed constraints \cite{cantor,molbiol}. In this complicated solution environment there are multiple ways that solute molecules can interact with one another and with the aqueous solvent.  The hydrogen-bonding (HB) network between water molecules, accounting for much of the anomalies present in its still contentious phase diagram \cite{pnas10}, stabilizes various distinct conformations of biopolymers providing short- and long-range interactions among the non-contiguous parts of the polymer chain. Obviously, there is a strong competition between the polymer-polymer and the polymer-water hydrogen bonding, which must be properly taken into account in order to describe the biopolymer conformational space or its modifications as a consequence of the action of co-solutes and solvents. In general a solute molecule can have a double effect on biopolymer conformations \cite{AdrianPNAS}: i) it can {\sl directly} bind to a biopolymer, therefore competing with intrinsic intra-polymer hydrogen bond configurations - we dub these {\sl competing interactions}, but also ii) it can {\sl indirectly} affect biopolymer conformation through changes in the activity of water, that then acts osmotically in tuning biopolymer configurations - we dub these {\sl non-competing interactions}. For different systems it is often possible to choose between the two frames of reference depending on the nature of the macromolecular system, the experimental design, and the properties that are being observed. 

Consistent with these different perspectives there are also a number of different theoretical approaches to biopolymer conformational changes, most notably the helix-coil transition, which will be the focus of our discussion here. Historically they have been quite often formulated in the context of {\sl spin models} and can thus creatively engage the whole repertoire of the theoretical methodology devised in that context \cite{flory,gros, mattice, goldstein,lr,bad10, stanprot, biopoly1, biopoly2,physa, ananik}. However, these different models are not equally conducive to a straightforward inclusion of solvent effects. For instance, the original Zimm-Bragg (ZB) model makes it difficult to account for microscopic details of the polymer-solvent interaction, and thus even a qualitative agreement with experiment is sometimes rather difficult to achieve \cite{peg}. In the ZB theory there are in fact two major parameters, the  {\sl stability}, $s$, and the {\sl cooperativity}, $\sigma$, that can be affected by the solvent \cite{faragopincus}. While the cooperativity parameter is assumed to be independent of the temperature, the stability parameter is temperature dependent. In fact it can be expressed through the free energy difference between the helix and coil conformations, $\Delta F$ as $s=\exp{[\Delta F / T]}=\exp{[\Delta U/T - \Delta S]}$; where $T$ is temperature, $U$ is the energy and $S$ the entropy \cite{polsher}, so that the transition temperature is determined from the condition $s=1$, implying a compensation between the energetic and entropic changes 
\cite{faragopincus}. 
This decoupling into the energy and entropy contributions is even more straightforward for other types of helix-coil transition models. In the Lifson-Roig (LR) approach the entropic and energetic parameters are in fact explicitly decoupled \cite{bad10,biopoly1,biopoly2}. The transition temperature is here obtained from the condition $\exp{[\Delta U/T]}=Q$, where again $\Delta U$ is the energy change at the transition and $Q$ the entropic penalty for the formation of hydrogen bonds.  In both approaches the solute molecule can affect biopolymer conformations through the dependence of the entropic and energetic parameters on the solvent properties.  Since the ZB and LR models can be alternatively formulated through the microscopic Hamiltonian of a more general type \cite{bad10,biopoly1, biopoly2}, dubbed the Generalized Model of Polypeptide Chain (GMPC), that also depends on an entropic parameter $Q$ and an energetic parameter $W$ ($=\exp{[\Delta U/T]}$),  the consideration of solvent effects on the level of the GMPC model level could be of paramount importance and can have far reaching consequences.

A major issue for the spin models of the aqueous solvent is to take proper account of the explicit tetrahedral HB network geometry, as well as the correct orientation of water molecules in close proximity to different molecular moieties along the biopolymer chain. Optimally the spin-like model of the solvent with hydrogen bonding ability should be rich enough to describe the specific intra-polymer H-bonding, as well as take into account the non-specific osmotic action of the solutes and thus exhaustively characterize the dual action of the solutes. As we show in what follows, the specific solvent-polymer H-bonding interaction inevitably redefines the temperature-dependent energetic parameter $W$ \cite{biopoly2,bad11} of the GMPC model, 
and on the other hand the non-specific type of interaction, as exemplified by polyethylene glycol (PEG)  \cite{peg,PRL,chris}, 
leads to the renormalization of the entropic parameter $Q$ (no direct analogue with ZB model, since both $s$ and $\sigma$ effectively include the entropy of coil $Q$), which becomes temperature-dependent. Obviously the equilibrium between the different biopolymer conformations can be altered by changing either $W$ or $Q$, so that both solvent mechanisms are relevant.

We will demonstrate that within the GMPC framework both mechanisms of solvent action can be dealt with on the same footing by tracing over the solvent degrees of freedom {\sl  explicitly} and {\sl exactly}, so that the ensuing effective partition function combines all the solvent effects in a unified framework with renormalized values of the parameters $W$ and $Q$. While this in itself is a major formal advance we also demonstrate how it can be used in the context of various problems involving conformational transitions of biopolymers in the aqueous solvent.


{The paper is organized as follows. We first summarize the solvent-free GMPC model of the helix-coil transition and briefly describe the methods we apply. Then, to account for both mechanisms of solvent action we complement the basic, \emph{in vacuo} Hamiltonian \cite{biopoly1,biopoly2} with two additive terms. Each of these terms has been treated separately before \cite{bad11,PRL}, but the detailed description of simultaneous effect of both has not been reported yet. In the Appendix we show how the solvent-related parts of Hamiltonian, that describe different mechanisms of action can be both traced over in the partition function to reduce the problem to the basic GMPC model with renormalized parameters. The proposed strategy allows us to generalize the problem of solvent description in such a way that both mechanisms of solvent-polymer interaction are described properly. The current paper is a logical extension of Ref. \cite{PRL} and complements the line of research presented in Refs. \cite{bad11, biopoly1,biopoly2}. Finally we show that many biopolymer properties and peculiarities in their behavior, such as the hot and cold denaturation, the temperature dependence of the hydrophobic effect and the unusual behavior of Intrinsically Disordered Proteins can be explained within the unified framework proposed in this work.}

\section{Solvent-free GMPC model}

The helical structure of biopolymers is stabilized mainly by intermolecular hydrogen bonding between repeat units; the presence of hydrogen bonds is a necessary prerequisite for the formation of the helix. Statistical description of the helix-coil transition requires three parameters: the energy parameter $W=V+1=\exp(U/T)$, where $U$ is the energy of the hydrogen bond; the entropy parameter $Q$, that stands for the ratio between the number of all accessible states versus the number of states available for the repeat unit in the helical conformation; and a geometric parameter $\Delta$, that describes the geometry of hydrogen bond formation. Hydrogen bond formation in polypeptides is known to affect three successive repeat units, thus $\Delta=3$ in any solvent. This parameter controls the size of the transfer matrix, dictating the transfer matrix size of the LR model to be $3 \times 3$ \cite{lr}. Instead, the other two parameters, $W$ and $Q$, can be altered by the presence of solvent. The Hamiltonian of the solvent-free GMPC model \cite{biopoly1,biopoly2} reads \cite{biopoly1}
\begin{equation}
\label{ham-basic}
-\beta H_0\left(\{\gamma_i\}\right)=J\sum\limits_{i=1}^{N}\delta _{i}^{(\Delta )}.
\end{equation}
\noindent Here $\beta=T^{-1}$, $N$ is the number of repeat units, and $J=U/T$ is the temperature-reduced energy of hydrogen bonding. We use a short-hand notation, \emph{e.g.}, $\delta_{j}^{(\Delta )}=\prod_{k=0}^{\Delta-1}\delta (\gamma_{j+k},1)$, where $\delta (x,1)$ stands for the Kronecker symbol and $\gamma_{l}=1,\ldots,Q$. The spin variable $\gamma$ describes the state of each repeat unit by assigning to each of them one of $Q$ possible conformations, number 1 corresponds to the helical conformation, and the remaining $Q-1$ to the coil conformations. In this way the important degeneracy of the coil state is taken into account. The partition function is then obtained as
\begin{equation}
\label{partfuncbasic}
\begin{gathered}
Z_{0}(V,Q) = \sum\limits_{\left\{ {\gamma _i=1} \right\}}^{Q} e^{-\beta H_0\left(\{\gamma_i\}\right)}=\sum\limits_{\left\{ {\gamma _i=1} \right\}}^{Q} \prod\limits_{i = 1}^N  \left[ {1+V\delta _i^{\left( \Delta  \right)}} \right].
\end{gathered}
\end{equation}
Alternatively, we may make use of the {\sl transfer-matrix formalism} and write
\begin{equation}
\label{partfuncbasictm}
\begin{gathered}
Z_{0}(V,Q) = \text{Trace } \hat{G}^N=\text{Trace } \hat{A} \ \hat{\Lambda}^N \ \hat{B}=\sum\limits_{k=1}^{\Delta}\lambda_{k}^{N},
\end{gathered}
\end{equation}
\noindent where
\begin{equation}
\label{tmatrix0}
\hat G(\Delta \times \Delta) = \left( {\begin{array}{*{20}c}
   {e^{J}} & 1 & 0 & {...} & 0 & 0 & 0  \\
   0 & 0 & 1 & {...} & 0 & 0 & 0  \\
   {...} & {...} & {...} & {...} & {...} & {...} & {...}  \\
   0 & 0 & 0 & {...} & 0 & 1 & 0  \\
   0 & 0 & 0 & {...} & 0 & 0 & {Q - 1}  \\
   1 & 1 & 1 & {...} & 1 & 1 & {Q - 1}  \\
 \end{array} } \right),
\end{equation}
\noindent with $\hat{A}$ and $\hat{B}$ being the corresponding left and right eigenvectors of matrix Eq.~\ref{tmatrix0}, while $\Lambda$ is the diagonal matrix of eigenvalues. The characteristic equation for solving the eigenvalue problem finally reads (see Ref.~\cite{biopoly1})
\begin{equation}
\label{chareq}
\lambda^{\Delta-1}(\lambda-W)(\lambda-Q)=(W-1)(Q-1).
\end{equation}
\noindent Its solution provides $\Delta$ eigenvalues $\lambda_k$ ($\lambda=-1$ eigenvalue has been added to write Eq.~\ref{chareq} in compact form and has no physical meaning). The form of Eq.~\ref{chareq} tells us that changes in both entropic parameter $Q$ and energetic parameter $W$ affect the equilibrium properties in the same way and, in principle, the same effect on polypeptide conformations can be achieved by changing either one of these parameters. 
\begin{figure}[!ht]
\begin{center}
\includegraphics[width=8.5cm]{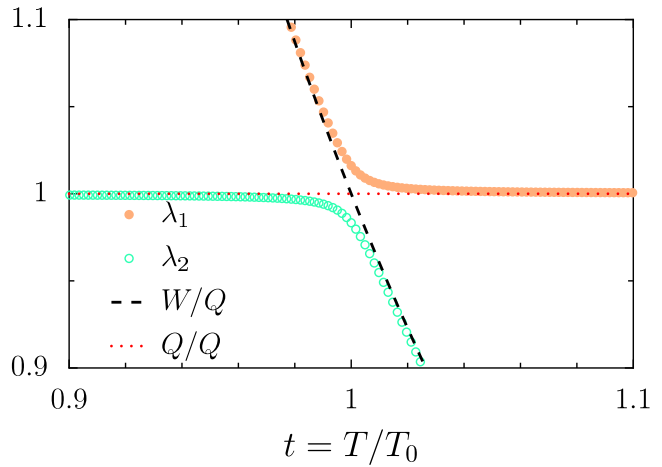}
\caption{\label{f0} The temperature dependence of two largest eigenvalues. The following set of parameters is used: $Q=60$, $\Delta=3$.}
\end{center}
\end{figure}
In the thermodynamic limit, the problem simplifies, it is enough to study the temperature dependence of the two largest eigenvalues of Eq.~\ref{chareq}. Eigenvalues come closest together at a point where the asymptotes $W(T)$ and $Q$ cross (Fig.~\ref{f0}). This is in accordance with the general physical considerations: the transition takes place at the point where entropy and energy compensate each other. The distance of minimal approach of eigenvalues can be estimated as $Q^{1-\Delta}$ (see \cite{biopoly2}) and is related to the final transition interval. Parameter $\Delta$ in our model plays the role of the spatial scale of the many-body interactions. For nearest neighbor interactions $\Delta=2$, for next-nearest neighbor interactions $\Delta=3$. We consider $\Delta=3$ for polypeptides and $\Delta=10$ for DNA, so $\Delta < \infty$ to mimic the short-range character of hydrogen bonding. 

If, instead, long-range interactions are assumed to act in the system, $\Delta \rightarrow \infty$ and the distance of minimal approach of eigenvalues would tend to zero so that the lowest two eigenvalues would be degenerate. In the spin language this signals the presence of a phase transition in the system. However, since we keep $\Delta$s finite, no phase transition {\sl sensu stricto} can happen in our model. To quantify the above, it is useful and informative to introduce the spatial correlation length as
\begin{equation}
\label{clength}
\xi=\ln^{-1}\left(\frac{\lambda_1}{\lambda_2}\right),
\end{equation}
\noindent where $\lambda_1$ and $\lambda_2$ are the first and second leading eigenvalues of the characteristic equation. Temperature - dependent $\xi$ has a maximum at the transition point. The height of the maximum is related to the transition interval as $\Delta T \sim \xi_{max}^{-1} \sim Q^{\frac{1-\Delta}{2}}$ \cite{biopoly1,biopoly2}. It turns out that the correlation function allows one to study both the stability and cooperativity of transition. Additionally, we can also easily calculate the degree of helicity as
\begin{equation}
\label{helicity}
\theta  = \left\langle {\delta _i^{(\Delta )} } \right\rangle  = \frac{1}
{N}\frac{{\partial \ln Z}}
{{\partial J}} = \frac{1}
{{\lambda _1 }}\frac{{\partial \lambda _1 }}
{{\partial J}}.
\end{equation}
Armed with this model and the methods of its solution we can proceed to generalize the original formulation by including the effects of the  solvents.

\section{Solvent effects within the GMPC model}

\subsection{Hydrogen bonding solvents (competing interactions)}

We assume the repeat units that are not bonded by intra-molecular H-bonds are free to form polymer-solvent intermolecular bonds and some solvents, such as water and urea, are able to form hydrogen bonds with nitrogen bases of DNA or with peptide groups of protein amino-acids \cite{molbiol,gros,cantor,war,she}. When one intra-molecular H-bond is broken, two solvent molecule binding sites become vacant. Thus, in the case of polypeptides, there are only two binding sites per repeat unit, while in the case of DNA there are four ($2 \times 2$ for an A-T pair) or six ($3 \times 2$ for an G-C pair) binding sites, so $2m$ $(m=1,2,3...)$ spin variables are required to describe the interaction between solvent molecules and each repeat unit. The reduced energy $J$ of the Hamiltonian in Eq.~(\ref{ham-basic}) now becomes $J=m\frac{(U_{pp}+U_{ss})}{T}$, where $U_{pp}$ and $U_{ss}$ are the energies of intra (polymer-polymer) and intermolecular (solvent-solvent) H-bonds, respectively. Such a model was considered in detail in Refs.~\onlinecite{biopoly1} and \onlinecite{bad11} using the following assumptions:
\begin{enumerate}
\item Only those repeat units of the polymer that do not participate in intermolecular hydrogen bonding are available for hydrogen bond formation with solvent.
\item Polymer-solvent interactions depend on the state (orientation) of solvent molecules with respect to the repeat unit;   there are $q$ possible discrete orientations of each solvent molecule.
\item A spin variable $\mu_i$, with values from $1$ to $q$, is assigned to each solvent molecule near repeat unit $i$. Orientation number $1$ is the bonded one, with energy $E$.
\item When intermolecular hydrogen bonding is broken in a polypeptide repeat unit, two binding sites become available. In the case of DNA there are two (A-T) or three (G-C) hydrogen bonds in one repeat unit, resulting in four or six binding sites available for solvent molecules. To generalize, we will consider $2m$ solvent spin variables per repeat unit. Here $m$ is the number of hydrogen bonds.
\end{enumerate}

The Hamiltonian for such a model of competing solvent (CS) reads
\begin{equation}
\label{hamcs}
-\beta H_{\text{CS}}(\{\gamma_i\},\{\mu_i^j\}) = I\sum\limits_{i = 1}^N {\left( {1 - \delta _i^{\left( \Delta  \right)} } \right) \cdot \sum\limits_{j = 1}^{2m} \delta \left( {\mu _i^j ,1} \right)},
\end{equation}
where $I=\frac{U_{ps}}{T}$ is the reduced energy of a polymer-solvent H-bond. Due to the presence of the term $1-\delta _i^{(\Delta)}$ in Eq.(\ref{hamcs}), as opposed to the $\delta _i^{(\Delta)}$ term in Eq.(\ref{ham-basic}), the solvent is competing with the polymer for H-bond formation, depending on the ratio $J/I$.

\subsection{Solvents affecting the available conformational space (non-competing interactions)}

There are many solvents or co-solutes that do not affect the hydrogen bonding directly, but do modify the polypeptide conformations by changing the chemical potential or the osmotic pressure of the solvent. A classical example of such a co-solute is PEG, which can act as an osmoticant and as a depletion agent \cite{Depletion}. Because of their size, PEG molecules are depleted from the proximal regions of the polypeptide chain, exerting an osmotic pressure that changes the energetic cost of certain conformations at the expense of others. We have introduced a model that describes these effects in \cite{PRL}. However, other types of solvents may exist besides the osmolytes. To cover all possible cases of non-H-bonding solvent, we model the solvent using the following assumptions.
\begin{enumerate}
\item Solvent can interact with (affect) both helical and coil units of polymer.
\item Interaction with a solvent molecule changes the energy of repeat unit depending on its conformation ($E_h$ if the repeat unit is helical and $E_c$ otherwise).
\item Polymer-solvent interaction depends on the orientation of the solvent molecule around the repeat unit; the number of solvent orientations being $p>2$ to account for solvent entropy.
\item A spin variable $\nu_i \in [1,p]$ is assigned to describe the state (orientation) of a solvent molecule and orientation number 1 is set to correspond to the case where binding takes place.
\end{enumerate}
The $\Delta E=E_h-E_c$ difference mimics the effect of the solvent. The larger this difference, the stronger is the stabilization of the helical state vs. the coil. Therefore we may qualitatively assume that $\Delta E$ models the effects of increased concentration of solvent. The corresponding Hamiltonian of non-competing solvent (NCS) reads
\begin{widetext}
\begin{equation}
\label{hamncs}
 - \beta H_{\text{NCS}}(\{\gamma_i\},\{\nu_i\}) = \sum\limits_{i = 1}^N {\left(I_c\left( {1 - \delta_i^{(1)} } \right) \delta \left( {\nu _i ,1} \right)+I_h\delta_i^{(1)} \delta \left( {\nu _i ,1} \right)\right)},
\end{equation}
\end{widetext}
\noindent where $I_{h,c}=E_{h,c}/T$.

\subsection{Solvent with combined dual interactions}

Interactions between some solvents (\emph{e.g.} urea), and a polymer have certainly an additional component besides the simple H-bonding. It may therefore happen that the same solvent affects polymer conformations through both effects: direct H-bonding and non-H-bonding mechanisms. Thus it seems to be a better idea to discuss the mechanisms of action and not specifically the solvent types. 

In general, the solvent can interact with the biopolymer by both mechanisms. In that case the general form of the Hamiltonian reads:
\begin{widetext}
\begin{equation}
\label{hamtotal}
 - \beta H_{\text{total}} = \sum\limits_{i = 1}^N \{
J\delta _i^{\left( \Delta  \right)}  +
I\left( {1 - \delta _i^{\left( \Delta  \right)} } \right) \sum\limits_{j = 1}^{2m} \delta \left( {\mu _i^j ,1} \right) +
I_c\left( {1 - \delta_i^{(1)} } \right) \delta \left( {\nu _i ,1} \right)+I_h\delta_i^{(1)} \delta ({\nu _i ,1}) \},
\end{equation}\end{widetext}
\noindent resulting in the partition function
\begin{equation}
\label{partfunct}
 Z_{\text{total}}= \sum\limits_{\{\gamma_i\}} \sum\limits_{\{\mu_i^j\}} \sum\limits_{\{\nu_i\}} \exp (-\beta H_{\text{total}}\left( \{\gamma_i\},\{\mu_i^j\},\{\nu_i\} \right)).
\end{equation}
\noindent Although the final Hamiltonian and the corresponding partition function look very complicated, all the solvent degrees of freedom can be analytically and explicitly summed out, without any assumptions, yielding the simple expression
\begin{equation}
\label{partfuncrenorm}
 Z_{\text{total}}= \left(q+e^{I}-1\right)^{2mN}\left(p+e^{I_h}-1\right)^{N}Z_{0}(e^{\widetilde{J}},\widetilde{Q}).
\end{equation}
\noindent Here
\begin{eqnarray}
\label{Wrenorm}
\widetilde{W} &=&\widetilde{V}+1=\exp{[\widetilde{J}]}=\exp{[\widetilde{U}/T]}=\nonumber\\
& &  \frac{q^{2m}e^J}{(q+e^{I}-1)^{2m}}=
\left( \frac{q \, e^{1/2t}}{q+e^{\frac{1+\alpha}{2t}}-1}\right ) ^{2m} 
\end{eqnarray}
\noindent and
\begin{equation}
\label{Qrenorm}
\widetilde{Q}=1+(Q-1) \frac{p+e^{I_c}-1}{p+e^{I_h}-1}=1+(Q-1) \frac{p+e^{\alpha_c/t}-1}{p+e^{\alpha_h/t}-1}.
\end{equation}
\noindent Above we have used the following notation: $t=2T/(U_{pp}+U_{ss})$, $\alpha=\frac{2U_{ps}-(U_{pp}+U_{ss})}{U_{pp}+U_{ss}}$, $\alpha_{h,c}=\frac{2E_{h,c}}{U_{pp}+U_{ss}}$. For the formal details of transformations that result in Eq.~\ref{partfuncrenorm}, see Appendix.

\section{results}

Eq.~\ref{partfuncrenorm} is the key result of our paper and means that the renormalization $W \longrightarrow \widetilde{W}$ and $Q \longrightarrow \widetilde{Q}$ in the transfer matrix \ref{tmatrix0}, the partition function \ref{partfuncbasictm} and the characteristic equation \ref{chareq} of the solvent-free GMPC model provides a full description of both types of solvent effects. To clarify the obtained results, it is informative to note that the characteristic equation that \emph{defines} the thermodynamics,
\begin{equation}
\label{chareqrenorm}
\lambda^{\Delta-1}(\lambda-\widetilde{W}(T))(\lambda-\widetilde{Q}(T))=(\widetilde{W}(T)-1)(\widetilde{Q}(T)-1),
\end{equation}
\noindent is similar to Eq.~\ref{chareq}. This means that even after the renormalization of the model parameters, the transition point (temperature) can still be determined from the intercept(s) between $\widetilde{W}$ and $\widetilde{Q}$. The two renormalized parameters with the changed temperature dependencies also lead to the changes in the phase diagram of the model. 

We first consider temperature dependencies of $\widetilde{W}$ and $\widetilde{Q}$ explicitly. From the their definitions it is clear that $W$ is exponentially decaying and that $Q$ is constant in the solvent-free model, as shown in Fig.~\ref{f0}.
\begin{figure}[!ht]
\begin{center}
\includegraphics[width=8.5cm]{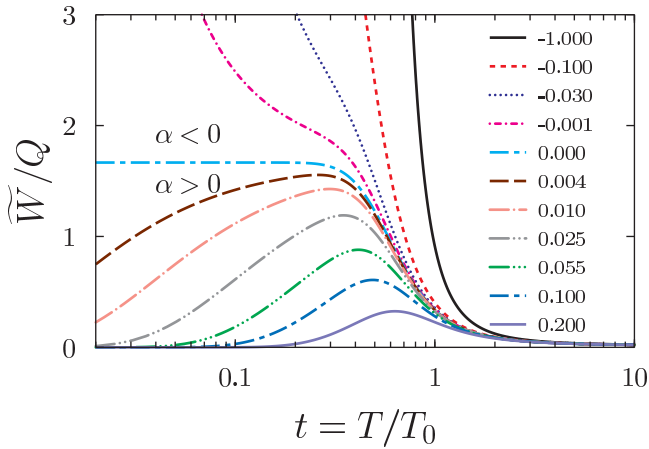}
\caption{\label{f1} Temperature dependence of $\widetilde{W}/Q$, with $Q=60$, $\Delta=3$. Curves are colored according to the values of $\alpha$.}
\end{center}
\end{figure}

\begin{figure}[!ht]
\begin{center}
\includegraphics[width=8.5cm]{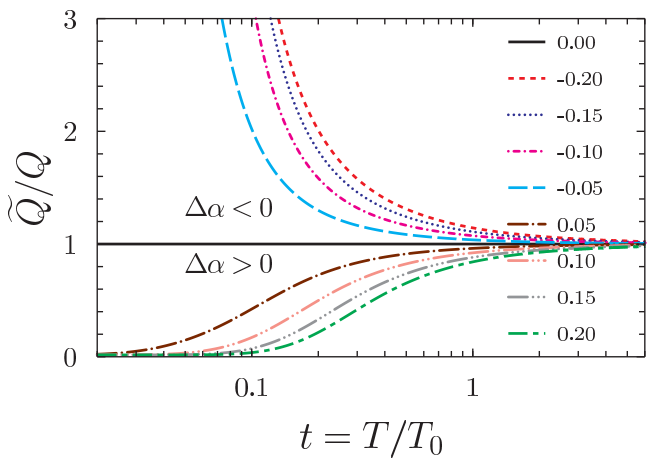}
\caption{\label{f2} Temperature dependence of $\widetilde{Q}/Q$, with $Q=60$, $\Delta=3$. Curves are colored according to the values of $\Delta\alpha$.}
\end{center}
\end{figure}
Figs.~\ref{f1} and \ref{f2} illustrate how the inclusion of solvent effects significantly changes the behavior of the model parameters which serve as asymptotes for the eigenvalues. To simplify these complicated dependencies, we will consider qualitatively different cases, controlled by energies of polymer-solvent interactions. 

There are three constants that tune the temperature behavior of $\widetilde{W}$ and $\widetilde{Q}$ and reflect the relative strength and sign of solvent-polymer interactions, namely, $\alpha=\frac{2U_{ps}-(U_{pp}+U_{ss})}{U_{pp}+U_{ss}}$ and $\alpha_{h,c}=\frac{2E_{h,c}}{U_{pp}+U_{ss}}$. For convenience, the last two constants will be further grouped into $\Delta \alpha=\alpha_h-\alpha_c=\frac{2(E_h-E_c)}{U_{pp}+U_{ss}}$. Combining representative curves from Figs.~\ref{f1} and \ref{f2} we can identify the transitions by looking for intercepts. As shown in Table \ref{tab1}, when describing the combined effect of the two mechanisms of solvent-polymer interactions, it is convenient to consider four possible cases:
\begin{center}
    \begin{tabular}{ | l | l | l |}
    \hline
                   & $\Delta \alpha < 0$ & $\Delta \alpha > 0$ \\ \hline
    $\alpha < 0$   & a)                  & b)  \\ \hline
    $\alpha > 0$   & c)  		 & d)   \\
    \hline
    \end{tabular}
\label{tab1}
\end{center}

\noindent Physically, these four cases correspond to situations when
\begin{enumerate}[(a)]
\item Polymer-polymer hydrogen bonding dominates, coil conformation is stabilized;
\item Polymer-polymer hydrogen bonding dominates, helical conformation is stabilized;
\item Polymer-solvent hydrogen bonding dominates, coil conformation is stabilized;
\item Polymer-solvent hydrogen bonding dominates, helical conformation is stabilized.
\end{enumerate}
Pure cases, when there is only one mechanism of action have been considered in our previous publications: $\Delta \alpha=0$ in Ref.~\cite{bad11} and $\alpha=-1$ in Ref.~\cite{PRL}. 

\begin{figure}[!ht]
\begin{center}
\includegraphics[width=8.5cm]{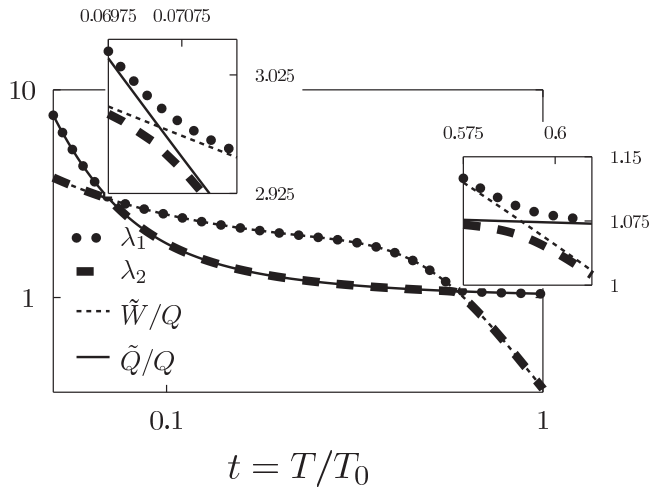}
\caption{\label{f3} {\bf Case a).} Temperature dependence of two largest eigenvalues of Eq.~\ref{chareqrenorm} at $\alpha<0$, $\Delta \alpha < 0$. The following set of parameters used: $Q=60$, $\Delta=3$, $\alpha=-0.01$, $\Delta \alpha =-0.03$.}
\end{center}
\end{figure}
\begin{figure}[!ht]
\begin{center}
\includegraphics[width=8.5cm]{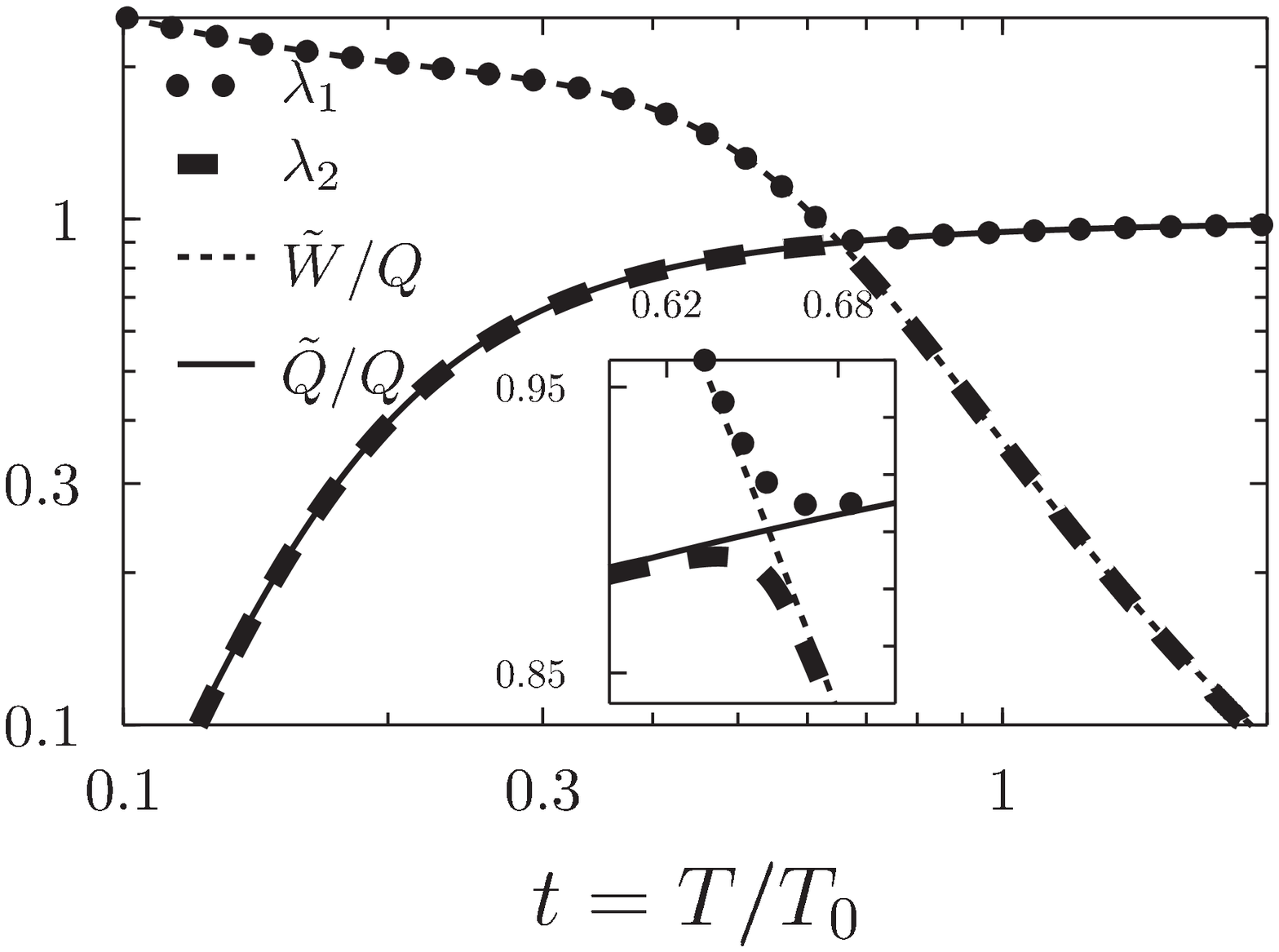}
\caption{\label{f4} {\bf Case b).} Temperature dependence of the two largest eigenvalues of Eq.~\ref{chareqrenorm} at $\alpha<0$, $\Delta \alpha > 0$; for $Q=60$, $\Delta=3$, $\alpha=-0.01$, $\Delta \alpha =0.03$.}
\end{center}
\end{figure}

\begin{figure}[!ht]
\begin{center}
\includegraphics[width=8.5cm]{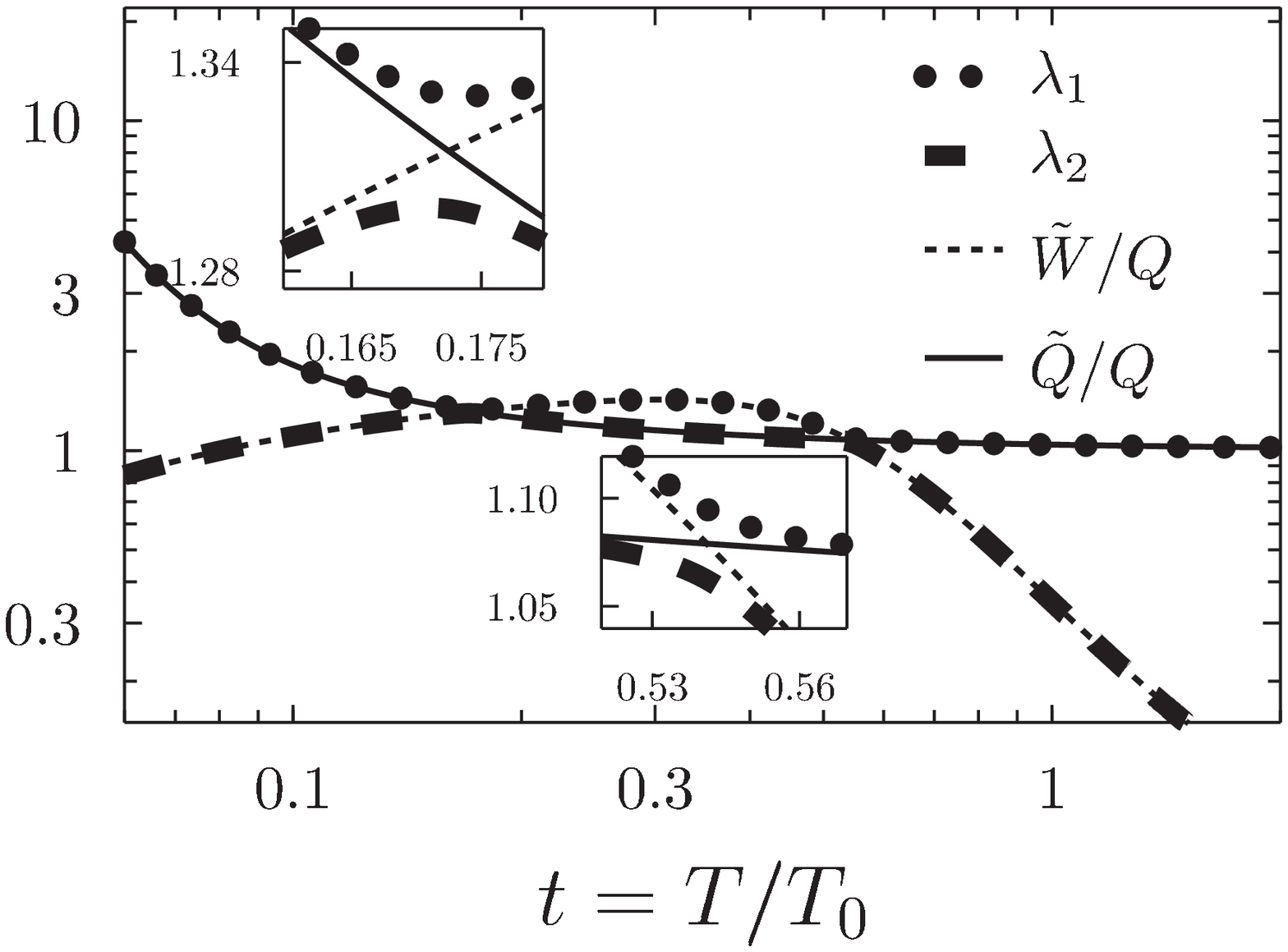}
\caption{\label{f5} {\bf Case c).} Temperature dependence of the two largest eigenvalues of Eq.~\ref{chareqrenorm} at $\alpha>0$, $\Delta \alpha < 0$; for $Q=60$, $\Delta=3$, $\alpha=0.01$, $\Delta \alpha =-0.03$.}
\end{center}
\end{figure}
\begin{figure}[!ht]
\begin{center}
\includegraphics[width=8.5cm]{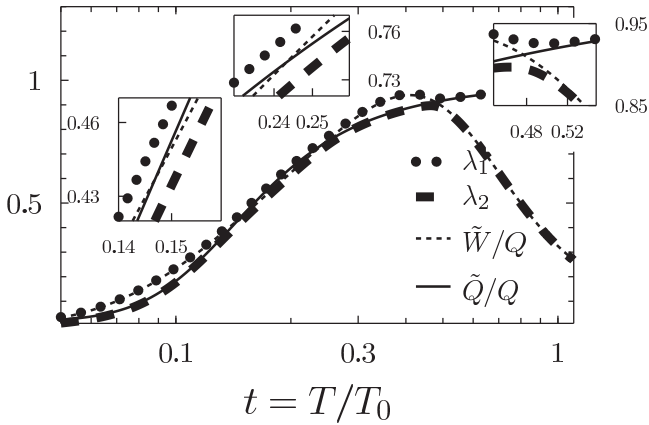}
\caption{\label{f6} {\bf Case d).} Temperature dependence of the two largest eigenvalues of Eq.~\ref{chareqrenorm} at $\alpha>0$, $\Delta \alpha > 0$; for $Q=60$, $\Delta=3$, $\alpha=0.014$, $\Delta \alpha =0.03$.}
\end{center}
\end{figure}
Representative plots for each of the four cases of Table ~\ref{tab1} are presented in Figs.~\ref{f3},\ref{f4},\ref{f5},\ref{f6}. As a direct consequence of main results, summarized by Eq.~\ref{partfuncrenorm}, even after redefinition, $\widetilde{W}$ and $\widetilde{Q}$ still remain to serve as asymptotes of two largest eigenvalues. These asymptotes intersect around the point of closest approach of eigenvalues.

At $\alpha<0$, in the absence of a non-competing solvent, there can only be one direct helix-coil transition (also see Ref.~\cite{bad11}). As we see in Fig.~\ref{f3}, the presence of destabilizing non-hydrogen bonding solvent with $\Delta \alpha < 0$ (case $a)$) on top of the destabilizing hydrogen bonding solvent, gives rise to an additional coil-helix transition at low temperatures. The reentrant transition wouldn't arise without the presence of the destabilizing non-hydrogen bonding solvent and corresponding low-temperature intercept appears due to the renormalization of $\widetilde{Q}$. 

At $\alpha<0$, $\Delta \alpha > 0$ (Fig.~\ref{f4}, case $b)$), additional stabilization of helical conformation by non-hydrogen bonding solvent (\emph{e.g.} PEG) does not qualitatively change the picture. There is only one, direct helix-coil transition at elevated temperature.

At $\alpha>0$, $\Delta \alpha < 0$ (Fig.~\ref{f5}, case $c)$) there are again two transitions. Polymer-solvent H-bonds dominate; addition of destabilizing non-hydrogen bonding solvent does not qualitatively change the situation. 

Finally, when $\alpha>0$, $\Delta \alpha > 0$ (Fig.~\ref{f6}, case $d)$), polymer-solvent H-bonding dominates, and non-competing solvent stabilizes the system. This is probably the most interesting case. It qualitatively corresponds to a water solution of polypeptides under the action of PEG osmotic stress. There are normally two transitions, but situations are possible, when another two transitions appear at lower temperatures. They are very unstable against small changes of $\alpha$ and $\Delta \alpha$. Such behavior for spin models is not unusual and has been reported before \cite{VWG}. The experimental observation of these low-temperature transitions is however difficult, since they will mostly appear at temperatures below the freezing point of water.

The overall behavior of the system is thus very rich, ranging from the case when there is no transition at all to the case when there are four transitions. For example, in case $d)$ at $\alpha=0.014$, $\Delta \alpha =0.03$ there are four transitions, at $\alpha=0.5$, $\Delta \alpha =1.0$ there are two transitions, while at $\alpha=0.5$, $\Delta \alpha =0.5$ there are no transitions at all and the system is always found in a disordered coil state. 

\begin{figure}[!ht]
\begin{center}
\includegraphics[width=8.5cm]{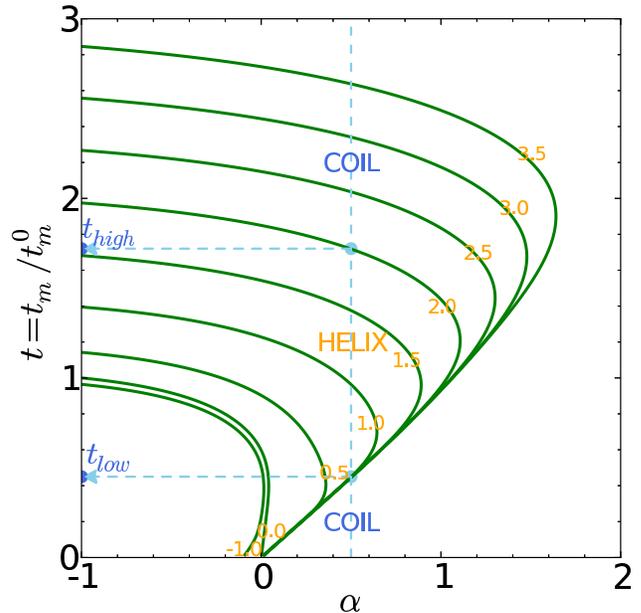}
\caption{\label{f7} Phase diagrams as temperature vs. $\alpha$ for some fixed $\Delta \alpha$ values shown on corresponding curves.}
\end{center}
\end{figure}

\begin{figure}[!ht]
\begin{center}
\includegraphics[width=8.5cm]{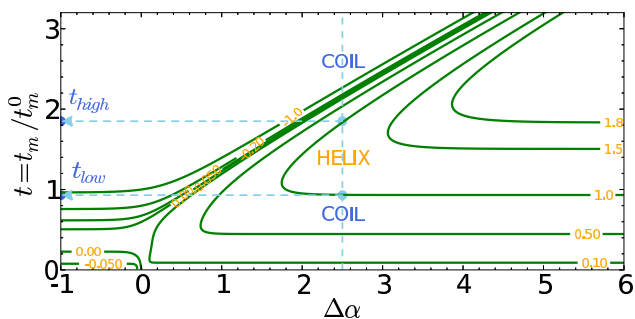}
\caption{\label{f8} Phase diagrams as temperature vs. $\Delta \alpha$ for some fixed $\alpha$ values shown on corresponding curves.}
\end{center}
\end{figure}

For better understanding of the situation we plot the "phase" diagrams. Namely, we wish to study how does the transition temperature change as a function of $\alpha$ for some fixed $\Delta \alpha$'s (Fig.~\ref{f7}) and as a function of $\Delta \alpha$ for some fixed $\alpha$'s (Fig.~\ref{f8}). The information from the curves can be extracted as follows. To withdraw the information for instance, from $\alpha=0.5, \\ \Delta \alpha=2.0$ case we draw the $\alpha=0.5$ line in Fig.~\ref{f7} (perpendicular dashed line) and look for its intercepts with the $\Delta \alpha=2.0$ curve. If there are intercepts, we project them onto the temperature axis (horizontal dashed lines) to find the transition temperatures (if any),  see Fig.~\ref{f8}.

The curves in Fig.~\ref{f7} indicate that at negative $\alpha$'s there is a transition from the helical to the coil conformation. At close to zero values of $\alpha$, more than one transition is possible; normally, there are two, but for small negative $\Delta \alpha$s situations are possible, when there are four transitions. The first transition point at low temperatures is from the coil to helix conformation and corresponds to the {\sl cold denaturation}, while the second, higher temperature point is for the regular transition from the helix to coil at elevated temperature. After some positive $\alpha$ there exists no transition point and the system will always be found in the coil conformation. This maximal value of $\alpha$ increases with increased $\Delta \alpha$. Thus the presence of non-competing solvent doesn't alter the phase diagram of polypeptides qualitatively, but can only shift the transition point to lower or higher temperatures.
Phase diagrams shown on Fig.~\ref{f8} look qualitatively different for the $\alpha>0$ and $\alpha<0$ case. For negative $\alpha$s there is always at least one, helix-coil ordinary transition. Transition point of the transition grows almost linearly with $\Delta \alpha$. At small negative or zero $\alpha$ the second branch may appear in the region of negative $\Delta \alpha$s of phase diagram, indicating the presence of a reentrant transition. For positive $\alpha$s phase diagrams are limited from the left and there exists a minimal value of $\Delta \alpha$, below which there is no transition at all, and above which there are two transitions. The transition temperature of ordinary helix-coil transition grows almost linearly with increased $\Delta \alpha$s at positive $\alpha$'s, like in the case of the negative $\alpha$'s. Interestingly enough, the transition point of the low temperature reentrant (coil-helix) transition is independent of $\Delta \alpha$. 

\begin{figure}[!ht]
\begin{center}
\includegraphics[width=8.5cm]{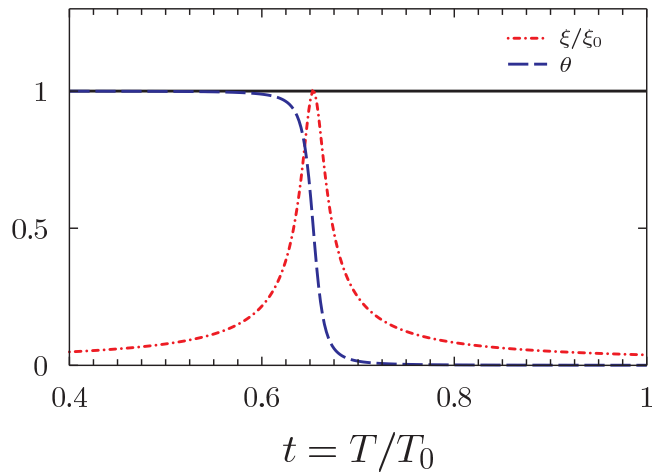}
\caption{\label{f10} Degree of helicity and correlation length (in reduced units) plotted at fixed $\alpha, \Delta \alpha$ (values shown in legend) indicate direct helix-coil transition.}
\end{center}
\end{figure}
\begin{figure}[!ht]
\begin{center}
\includegraphics[width=8.5cm]{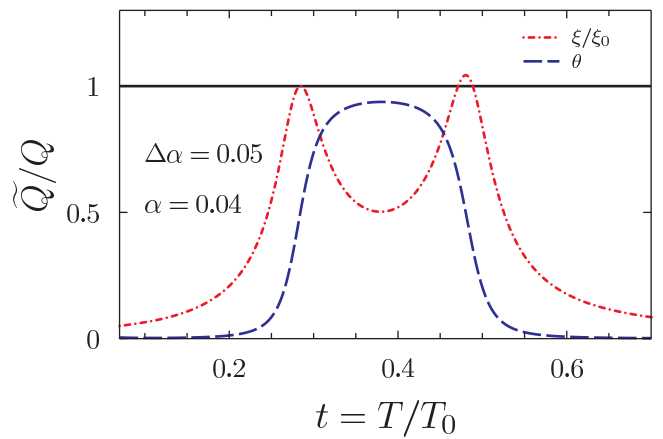}
\caption{\label{f9} Degree of helicity and correlation length (in reduced units) plotted at fixed $\alpha, \Delta \alpha$ (values shown in legend) indicate reentrant coil-helix at low temperatures followed by direct helix-coil transition at higher temperatures.
}
\end{center}
\end{figure}

The cases considered can be additionally visualized with the help of temperature dependencies of the degree of helicity $\theta$ and the spatial correlation length $\xi$. If only a direct, helix-coil transition is present, there will be one step of the helicity degree and only one peak in the correlation length (Fig.~\ref{f10}), while if there is an additional, reentrant transition, two steps and two peaks appear (Fig.~\ref{f9}).
An important result follows from Fig.~\ref{f9}. While in the absence of non-hydrogen bonding solvent the maxima of the correlation lengths are equal for both the reentrant and direct transitions (see Fig. 2 of Ref. \cite{bad11}), resulting in similar cooperativities and transition intervals, experimental results indicate that the cooperativities and intervals of heat and cold denaturations do differ \cite{privalov}. In the language of the correlation length this means different values of maxima, see Fig.~\ref{f9}. Depending on the signs and values of $\alpha$ and $\Delta \alpha$ it could even happen that the maximum at low temperatures is larger or smaller than the high temperature one. 

\begin{figure}[!ht]
\begin{center}
\includegraphics[width=8.5cm]{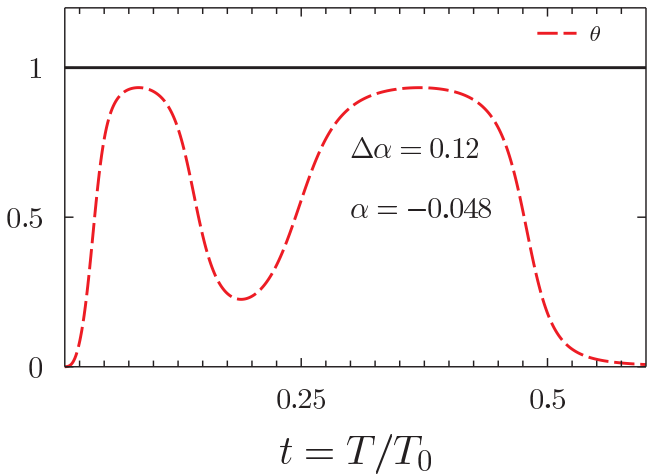}
\caption{\label{f11} The degree of helicity plotted for $\alpha=0.048$ and $\Delta \alpha=0.12$ indicates the possibility of four transitions.}
\end{center}
\end{figure}

Besides the simple cases shown above, more complex situations are possible, including the case of four transitions (Fig.~\ref{f11}) for which the helical content does not reach one or zero.
Cases when the helical content doesn't reach saturation are potentially interesting for studies of Intrinsically Disordered Proteins (IDP), which normally have a low number of secondary structure elements. Effectively, it means that IDPs have lower rigidity as compared to "normal", ordered proteins, and therefore, cannot fold at conditions when other proteins are folded. Our results allow us to explain many regularities in the behavior of IDPs and to determine their place in the general phase diagram. It appears, that IDPs at room temperature are in the region below cold denaturation point, so that they gain order upon heating; for some of them, however, the normal unfolding transition is preempted by the water boiling point.

\section{conclusion}

Due to the complex character of interactions between the biopolymers and the aqueous solvent, with few advances in the theory of HB liquids (because of the absence of small parameter), finding and linking together proper models for both the solvent and the polymer is not a simple task. Since directional HB interactions play the most important role in the system, it seems natural to rely on spin models which abound in the literature. The coupled Ising-Potts model \cite{VWG} was quite successful in describing the lower critical solution points in hydrogen-bonded mixtures. The Potts spin framework has been implemented successfully in order to describe the cold and the warm swelling of hydrophobic polymers in water \cite{delosrios1}, as well as the general case of chaotropic and hydrophobic solvents \cite{delosrios2}. The nature and description of the solvent itself can be a major issue since the water phase diagram can be quite baroque with unusual phase structure possibly involving a second critical point \cite{stanley_group}. However, even these exotic scenarios could be modeled within the spin model Hamiltonian, e.g. in the context of a Bell-Lavis spin model, that allows for a reentrant phase diagram involving low and high density phases of water as was recently pointed out \cite{fiore09}. 

However, because we are mostly interested in the influence of the solvent on the biopolymer conformations and not vice versa, it would seem that a detailed description of the bulk solvent is of a lesser relevance \cite{bad11,PRL}. Excluding the extremophiles, biological systems thrive at temperatures between the freezing point and the boiling point of water. Therefore, there is no real need to describe the critical properties of solvents in this context. On the other hand, water is actively rearranging its H-bonding network even at (and below) room temperature, so that it cannot be described as some solid and unresponsive medium. Additionally, the polymer-solvent interactions that we take into account are short ranged in space and thus allow to significantly simplify the description of solvent-induced effects on the polymer and to reduce the solvent description from a three-dimensional one to a one-dimensional one. 

By considering two separate models of solvents we described two different \emph{mechanisms of solvent-biopolymer interaction},  corresponding to explicit and implicit interactions, simultaneously and on the same footing. For instance, PEG molecules of intermediate length are big enough to act implicitly as osmolytes creating osmotic stress, while at the same time the hydrogen bonding ability of low molecular ethylene glycol could still be affecting the interactions explicitly. For PEG molecules of $\sim 100$ repeat units the hydrogen bonding activity could be disregarded, since fluctuations will destroy any direct H-bonds between large PEG molecules and the polypeptides, but for shorter PEG molecules both mechanisms can play a role. In general, it becomes an interesting and still open question what is the most important mechanism of action for solvents like urea and guanidine in solution with water and polypeptides. Is cold denaturation in solutions with urea or guanidine arising due to increased preference of polymer-solvent hydrogen bonding (increased $\alpha$), or is it due to osmotic stabilization (increased $\Delta \alpha$), or in fact both? 

The possibility of having both heat and cold denaturation is a property that results from the directionality of H-bonding interactions and this is a feature shared by both polypeptides and polynucleotides. Often, however, cold denaturation temperature appears at temperatures below the water freezing point, making the experimental observation impossible. The situation for DNA is even worse. In fact the smaller the entropy of the coil state, the lower is the reentrant transition temperature (see Fig. 3 of of Ref. \cite{bad11}), so that the cold denaturation of DNA would be very difficult to observe. These facts make our theoretical considerations of great importance, since adding a non-competing solvent to the solution may potentially make the observation of cold denaturation possible even for systems where such observations would be difficult otherwise. 

While skepticism has been voiced in the literature that an implicit description of the solvent is unlikely to account for for both cold and heat denaturation unless the model parameters are fitted to thermodynamic properties (e.g., temperature-dependent energetics) \cite{Matysiak}, our implicit models of solvent do exactly what has been deemed as "unlikely". Also the conviction that only explicit solvent models can describe the heat and cold denaturation naturally from first principles \cite{Matysiak} does not seem to so self-evident as it would appear from the example of e.g. the popular Mercedes-Benz (MB) model of water \cite{bennaim,dill,kartun1,kartun2}. In fact our model of a solvent with H-bonding interactions is conceptually  very close to the explicit water models with directional interactions, like the MB model. Indeed, the most natural analytic way to describe the orientational interactions on the Hamiltonian level is through the multivalued spin variables \cite{baxter}, like what we did for the GMPC model. It might thus be appropriate to rather adjust the strict statement on the explicit solvent models into a softer statement that the proper model of the solvent need to exhibit explicitly only the \emph{directional interaction} of the solvent in order to recover both the heat and the cold denaturation naturally from first principles.

Another potentially fruitful research area where the proposed theory may be of great importance, is the "unusual" behavior of Intrinsically Disordered Proteins. First, although disordered, they are functional and resistant to cold treatment \cite{tantos}, so we are still missing some crucial info about the basics of the folding event itself. Many IDP's also gain structure upon increasing temperature in the range from 3 to 50 degrees Celsius or decreasing pH from 5.5 to 3.0 \cite{uversky1}. Also of note is that some IDP's in crowded environment are folding, while others remain unfolded \cite{uversky2}. These facts, coupled with our results, summarized in phase diagrams Figs.~\ref{f7} and \ref{f8}, show that the disordered state of IDP's belongs to the low-temperature region of the phase diagram, so that decreasing temperature doesn't have any effect, while increasing it results in cold denaturation and/or refolding. This transition takes place at conditions where globular proteins usually loose structure, while IDP's gain it. Resistance to crowding is nicely visible in Fig.~\ref{f8}, where the low-temperature part of the curve describing cold denaturation is almost parallel to the $x$-axis that describes the crowding in the system ($\Delta \alpha$). In the same figure it is also visible that the more pronounced is the competition between the inter- and the intra-molecular hydrogen bonding (larger $\alpha$ values), the higher is the temperature of cold denaturation, which explains why some IDP's gain order in crowding conditions while others do not.

The present study of solvent effects has important implications on both polypeptides and DNA and the qualitative picture that we derived is very reach, including the possibility for both the reentrant as well as direct helix-coil transitions, enabling situations when only a certain amount of helicity is lost/gained. Such effect might share light on changes of disordered protein conformations and DNA replication and explain how these processes are regulated by solvents inside a cell. Our theory also allows for explanations of the unusual behavior of the Intrinsically Disordered Proteins, thus showing a strong potential for future studies in this vastly developing research field. While we have limited ourselves to the helix-coil transition phenomenon, the approach advocated is extendable to any spin-based theory of conformational transitions in polymers. 

\section{acknowledgment}

RP acknowledges support from the Agency for Research and Development of Slovenia (ARRS grants No. J1-4297 and J1-4134). YM and VM acknowledge support from the State Committee of Science of the Republic of Armenia (grant No. 13-1F343) and from the Volkswagen Foundation (grant "Equilibrium and non–equilibrium behavior of single– and double–stranded biological molecules").

\appendix
\section{Exact integration of the solvent degrees of freedom}
Eq.~(\ref{partfunct}) reads
\begin{equation}
\label{partfunct1}
 Z_{\text{total}}= \sum\limits_{\left\{ {\gamma _i } \right\}} \prod\limits_{i = 1}^N \left[ {1+V\delta _i^{\left( \Delta  \right)}} \right] \times L_{\text{CS}}({\left\{ {\gamma _i } \right\}}) \times M_{\text{NCS}}({\left\{ {\gamma _i } \right\}}),
\end{equation}
\noindent where
\begin{widetext}
\begin{equation}
\label{L_i}
\begin{gathered}
L_{\text{CS}}({\left\{ {\gamma _i } \right\}}) \equiv \sum\limits_{\left\{ {\mu_i^j } \right\}} \prod\limits_{j = 1}^{2m} \left[ {1+R(1-\delta _i^{\left( \Delta  \right)}}) \cdot \delta \left( {\mu _i^j ,1} \right) \right]= \\ \sum\limits_{\mu_i^1 =1}^q \sum\limits_{\mu_i^2 =1}^q ... \sum\limits_{\mu_i^{2m} =1}^q  \{ 1+R(1-\delta _i^{\left( \Delta  \right)}) \sum\limits_{j=1}^{2m} \delta \left( {\mu _i^j ,1} \right)+ R^2(1-\delta _i^{\left( \Delta  \right)}) \sum\limits_{j<k}\delta \left( {\mu _i^j ,1} \right)\delta \left( {\mu _i^k ,1} \right)+
 \\ R^3(1-\delta _i^{\left( \Delta  \right)}) \sum\limits_{j<k<l}\delta \left( {\mu _i^j ,1} \right)\delta \left( {\mu _i^k ,1} \right)\delta \left( {\mu _i^l ,1} \right)+... \\ R^{2m}(1-\delta _i^{\left( \Delta  \right)})\delta \cdot \left( {\mu _i^1 ,1} \right) \cdot \delta \left( {\mu _i^2 ,1} \right)...\cdot \delta \left( {\mu _i^{2m} ,1} \right) \} = \\ q^{2m}+(1-\delta _i^{\left( \Delta  \right)})\left[2mRq^{2m-1}+C_{2m}^2 R^2 q^{2m-2}+C_{2m}^3 R^3 q^{2m-3}+...+R^{2m}\right]= \\q^{2m}+(1-\delta _i^{\left( \Delta  \right)})\left[q+R\right]^{2m}-(1-\delta _i^{\left( \Delta  \right)})q^{2m}= (q+R)^{2m} \left[ 1-\delta _i^{\left( \Delta  \right)}+\frac{q^{2m} \delta _i^{\left( \Delta  \right)}}{(q+R)^{2m}}\right].
\end{gathered}
\end{equation}
\end{widetext}
$V=e^{J}-1$ and $R=e^{I}-1$ have been introduced as in the main text. Above we have just summed out degrees of freedom of competing solvent, used the properties of the binomial coefficients $C_{n}^{m}=n!/(m!(n-m)!)$ and rearranged the terms. 
In its turn, $M_{\text{NCS}}$ can be simplified without assumptions too:
\begin{widetext}
\begin{equation}
\label{M_i}
\begin{gathered}
M_{\text{NCS}}({\left\{ {\gamma _i } \right\}}) \equiv \sum\limits_{\left\{ {\nu_i} \right\}} \left[ {1+R_c(1-\delta _i^{\left( 1  \right)}})\delta \left( {\nu _i ,1} \right) \right] \left[ {1+R_h\delta _i^{\left( 1  \right)}}\delta \left( {\nu _i ,1} \right) \right]= (p+R_c)(1+\rho \delta _i^{\left( 1  \right)}),
\end{gathered}
\end{equation}
\end{widetext}
\noindent where $\rho=\frac{R_h-R_c}{p+R_c}$.

Inserting $L_{\text{CS}}$ into Eq.(\ref{partfunct1}) results in:
\begin{widetext}
\begin{equation}
\label{partfunct2}
 Z_{\text{total}}= (q+R)^{2mN}\sum\limits_{\left\{ {\gamma _i } \right\}} \prod\limits_{i = 1}^N \left[ {1+\widetilde{V}\delta _i^{\left( \Delta  \right)}} \right] \times M_{\text{NCS}}({\left\{ {\gamma _i } \right\}}),
\end{equation}
\end{widetext}
\noindent where
\begin{equation}
\label{Vtilde}
\widetilde{V}+1=\exp{(\widetilde{J})}=\frac{(V+1)q^{2m}}{(q+R)^{2m}}=\frac{\exp{(m\frac{(U_{pp}+U_{ss})}{T})}\cdot q^{2m}}{\left[ q-1+\exp{(\frac{U_{ps}}{T})}\right]^{2m}}.
\end{equation}
\noindent

Further insertion of $M_{\text{non-hbsol}}({\left\{ {\gamma _i } \right\}})$ and summation over non competing solvent degrees of freedom gives us:

\begin{equation}
\label{partfunct3}
 Z_{\text{total}}= (q+R)^{2mN}(p+R_h)^{N}\sum\limits_{\left\{ {\gamma _i } \right\}} \prod\limits_{i = 1}^N \left[ {A_i+\widetilde{V}\delta _i^{\left( \Delta  \right)}} \right].
\end{equation}

The last expression is up to (unimportant) constants similar Eq.~\ref{partfuncbasic} of solvent-free case, with the only difference that the function to be summed contains 
\begin{equation}
\label{ai}
A_i=\frac{1+\rho\delta _i^{\left( 1  \right)}}{1+\rho},
\end{equation}
\noindent on the place of $1$. The product inside Eq.~\ref{partfunct3} results in polynomial of degree $N$ of parameter $V$ with coefficients, that contain different combinations of $A_i$. Summation over all $\gamma_i$s of each of terms of this polynomial gives us the partition function of the model with solvent. In solvent-free case we had $\sum\limits_{\gamma _k=1}^Q 1=Q$ which is now $\sum\limits_{\gamma _k=1}^Q A_k=\frac{Q+\rho}{1+\rho}=1+(Q-1)\frac{p+R_c}{p+R_h}=\widetilde{Q}$. By changing the summation limit, $\sum\limits_{\gamma _k=1}^{\widetilde{Q}} 1=\widetilde{Q}$ Eq.(\ref{partfuncrenorm}) is obtained.

\bibliographystyle{apsrev}

\end{document}